\documentclass[12pt]{article}

\usepackage{slashed}
\usepackage{latexsym,epsfig,amssymb}
\usepackage{amsmath}
\usepackage{slashed}
\usepackage{mathrsfs}
\usepackage[bookmarks=false,colorlinks=true,linkcolor=blue,citecolor=blue,linktoc=page]{hyperref}

\begin{document}

\thispagestyle{empty}

\begin{flushright}\small
MIFPA-12-13, YITP-SB-12-08
\end{flushright}

\bigskip
\bigskip

\vskip 5mm
\begin{center}
  {\Large {\bf{New superconformal models in six dimensions:}}}\\\vspace{2mm}

       {\Large {\bf{Gauge group and representation structure}\footnote{Proceedings of the Corfu Summer Institute 2011 "School and Workshops on Elementary Particle Physics and Gravity",
		 September 4-18, 2011,
		 Corfu, Greece.}}}\\[3ex]
\end{center}
\vskip 3mm
\begin{center}
{\bf Henning Samtleben$^{a}$, Ergin Sezgin$^b$, Robert Wimmer$^{a,c}$, Linus Wulff$^b$}\\[3ex]
\end{center}
\vskip 3mm
\noindent
$^a$\,{\em Universit\'e de Lyon, Laboratoire de Physique,\\ UMR 5672, CNRS et ENS de Lyon,\\
46 all\'ee d'Italie, F-69364 Lyon CEDEX 07, France} \\
\vskip 3mm

\noindent
$^b$\,{\em George P. and Cynthia W. Mitchell Institute \\for Fundamental
Physics and Astronomy \\
Texas A\&M University, College Station, TX 77843-4242, USA}\\
\vskip 3mm

\noindent
$^c$\,{\em C.N. Yang Institute for Theoretical Physics\\
       State University of New York\\
       Stony Brook, NY 11794-3840, USA.}\\

\vskip1.8cm
\begin{center} {\bf Abstract } \end{center}
\begin{quotation}\noindent
We review recent progress \cite{Samtleben:2011fj} in the
construction and classification of six-dimensional
(1,0) superconformal models with non-abelian tensor fields.
Here we solve the generalized Jacobi identities which are required for
consistency of the non-abelian vector/tensor gauge system and
we present a large class of explicit examples.
\end{quotation}

\vfill

\noindent
{\footnotesize{henning.samtleben@ens-lyon.fr, sezgin@physics.tamu.edu,\\ wimmer@insti.physics.sunysb.edu,
linus@physics.tamu.edu}}


\newpage
\setcounter{page}{1}


\section{Introduction}


The understanding of six-dimensional superconformal theories in the context of multiple M$5$-branes is
one of the  pressing questions in M-theory. Already the mere existence of such higher dimensional non-trivial
quantum field theories is an exciting result \cite{Witten:1995zh}. However, in general it is believed that
an explicit understanding of such theories is out of reach.

One problem in this context is the description of non-abelian tensor (two-form) gauge fields. For example,  a no-go theorem
states that there exists no non-abelian extension of the abelian tensor gauge symmetry \cite{Bekaert:2000qx}.
In \cite{Samtleben:2011fj} it was shown how to circumvent this problem in the context of a tensor hierarchy \cite{deWit:2005hv}
by introducing additional form-degrees of freedom, in particular a non-propagating three-form.
The other problem is that the supposed $(2,0)$ theory of multiple M$5$-branes is intrinsically strongly
coupled, i.e.\ it has no free parameter for a weak coupling expansion which would make the
existence of a Lagrangian description or equations of motions plausible. This problem is analogous to the situation of M$2$-branes. Also
in this case, and for the same reason, it was believed that a Lagrangian description does not exist. Nevertheless,
for the isolated case of an $SO(4)$ gauge group a Lagrangian description
of a maximally supersymmetric three-dimensional Chern-Simons-matter CFT (BLG-model) has been found
\cite{Bagger:2007jr,Gustavsson:2007vu}. Subsequently, a more general class of Lagrangian three-dimensional CFT's,
with  gauge groups of arbitrary rank, has been formulated in \cite{Aharony:2008ug} (ABJM-models). The decisive observation in the latter
case is that by placing the M$2$-branes at an orbifold singularity instead of placing them in flat space one gains a dimensionless
parameter which allows for a weak coupling limit and thus makes a Lagrangian description possible.
The resulting CFT's have the same field content as a maximally, i.e. ${\cal N}=8$ supersymmetric theory but
realize only ${\cal N}=6$ supersymmetry. From the field theory point of view this means that the reduced supersymmetry is
less restrictive and therefore allows for a Lagrangian formulation.

The present article is based on the findings of \cite{Samtleben:2011fj}whose the main idea
is in the spirit of the ABJM-models: Instead
of considering the maximal supersymmetric case with ${\cal N}=(2,0)$ supersymmetry one considers
${\cal N}=(1,0)$ supersymmetry, with eventually the same field content as the ${\cal N}=(2,0)$ theories. This led to
an explicit dynamical description of superconformal non-abelian tensor multiplets through equations of motions or Lagrangians,
respectively. Crucial in the implementation of a non-abelian gauge and tensor gauge symmetry was the construction
of a tensor hierarchy of one-, two- and (non-dynamical) three-forms. The tensor hierarchy is formulated in terms of a number of
invariant tensors of the to be determined structure group. Consistency of this
gauge symmetry imposes several non-linear conditions on these invariant tensors, which represent
{\em generalized Jacobi identities}. These conditions play an essential role also in deriving the dynamics from
supersymmetry. Generically the dynamics is described by equations of motions only, but for a subclass also a
Lagrangian formulation exists.
Models defined by equations of motions but without a proper Lagrangian description
might be still of interest and give upon dimensional reduction to five dimensions a Lagrangian dynamics
with the correct inverse compactification radius behavior \cite{Witten:2007ct}.

In the following we review the construction of \cite{Samtleben:2011fj} and in particular, we present the general
solutions to the
generalized Jacobi identities for the case of semi-simple groups. We also discuss the resulting classes of
${\cal N}=(1,0)$ superconformal models of non-abelian tensor multiplets.


\section{Non-abelian tensor fields in six dimensions}
\label{sec:nonab}



\subsection{Review of general non-abelian vector/tensor systems}
\label{subsec:nonab}


In this section, we review the results of~\cite{Samtleben:2011fj} on the general structure
of non-abelian vector/tensor systems in six dimensions, based on the non-abelian $p$-form tensor hierarchy, as
worked out in~\cite{deWit:2005hv,deWit:2008ta,Bergshoeff:2009ph,Hartong:2009vc}.
We will be discussing gauge theories with field content given by the $p$-forms
$\{ A_\mu^r, B_{\mu\nu}{}^I, C_{\mu\nu\rho\,r}\}$, where the indices $r$ and $I$ label
the vector and the two-form tensors, respectively.
Anticipating six-dimensional dynamics with vector fields dual to antisymmetric three-form tensors,
we use a dual internal index $r$ to label the latter fields.

For vector and two-form tensor fields, the full covariant
non-abelian field strengths are given by
\begin{eqnarray}
{\cal F}_{\mu\nu}^r &\equiv&
2 \partial_{[\mu} A_{\nu]}^r - f_{st}{}^r A_\mu^s A_\nu^t + h^r{}_I\,B_{\mu\nu}^I
\;,\nonumber\\[.5ex]
{\cal H}_{\mu\nu\rho}^I &\equiv& 3 D_{[\mu} B_{\nu\rho]}^I +
6 \, d^I{}_{rs}\,  A_{[\mu}^r \partial^{\vphantom{r}}_\nu A_{\rho]}^s
- 2 f_{pq}{}^s d^I{}_{rs}\, A_{[\mu}^r A_\nu^p A_{\rho]}^q
+ g^{Ir} C_{\mu\nu\rho\,r}
\;,
\label{defF}
\end{eqnarray}
in terms of the antisymmetric structure constants $f_{st}{}^r=f_{[st]}{}^r$,
a symmetric $d$-symbol $d^I{}_{rs}=d^I{}_{(rs)}$, and the tensors $g^{Ir}$, $h^r{}_I$
inducing general St\"uckelberg-type couplings among forms of different degree.\footnote{
We use canonical dimensions such that a $p$-form has mass dimension $p$
and as a result all constant tensors $f_{st}{}^r$,
$d^I{}_{rs}$, $g^{Ir}$, $h^r{}_I$, are dimensionless.
}
The covariant derivatives are defined as $D_\mu \equiv \partial_\mu - A_\mu^r X_r$ with
an action of the gauge generators $X_r$ on the different fields
given by $X_r \cdot \Lambda^s \equiv - (X_{r})_{t}{}^s \Lambda^t$,
$X_r \cdot \Lambda^I \equiv - (X_{r})_{J}{}^I \Lambda^J$, etc.
The field strengths (\ref{defF}) are defined such that they transform covariantly under the
set of non-abelian gauge transformations
\begin{eqnarray}
\delta A_\mu^r &=& D_\mu \Lambda^r - h^r{}_I \Lambda_\mu^I
\;,\nonumber\\[.5ex]
\Delta B_{\mu\nu}^I &=& 2 D_{[\mu} \Lambda_{\nu]}^I -2\, d^I{}_{rs} \,\Lambda^r {\cal F}_{\mu\nu}^s
- g^{Ir} \Lambda_{\mu\nu\,r}
\;,\nonumber\\[.5ex]
\Delta C_{\mu\nu\rho\,r} &=& 3 D_{[\mu} \Lambda_{\nu\rho]\,r}
+3 \, b_{Irs}\,{\cal F}_{[\mu\nu}^s \,\Lambda_{\rho]}^I
+ b_{Irs}\,{\cal H}_{\mu\nu\rho}^I \,\Lambda^s
+\dots
\;,\label{gaugesym}
\end{eqnarray}
where we have introduced the compact notation
\begin{eqnarray}
\Delta B^I_{\mu\nu} &\equiv& \delta B^I_{\mu\nu} - 2 d^I{}_{rs}\,A_{[\mu}^r \,\delta A_{\nu]}^s
\;,\nonumber\\
\Delta C_{\mu\nu\rho\,r} &\equiv& \delta  C_{\mu\nu\rho\,r}
-3\, b_{Irs}\,B_{[\mu\nu}^I \,\delta A_{\rho]}^s
-2\, b_{Irs}\, d^I{}_{pq}\,A_{[\mu}^s \,A_{\nu}^p \,\delta A_{\rho]}^q
\;.\label{Delta1}
\end{eqnarray}
The ellipsis in the last line of (\ref{gaugesym}) represent possible terms that
vanish under projection with $g^{Ir}$.
The covariant field strengths (\ref{defF}) satisfy the modified
Bianchi identities
\begin{eqnarray}
D^{\vphantom{r}}_{[\mu} {\cal F}_{\nu\rho]}^r &=& \frac13 h^r{}_I\,{\cal H}_{\mu\nu\rho}^I
\;,\nonumber\\[.5ex]
D^{\vphantom{I}}_{[\mu} {\cal H}_{\nu\rho\sigma]}^I &=&
\frac32 d^I{}_{rs} \, {\cal F}^r_{[\mu\nu} {\cal F}^s_{\rho\sigma]}+
\frac14 g^{Ir} {\cal H}^{(4)}_{\mu\nu\rho\sigma\,r}
\;,
\label{Bianchi}
\end{eqnarray}
where the non-abelian field strength ${\cal H}^{(4)}_{\mu\nu\rho\sigma\,r}$
of the three-form potential is defined by the second equation.
In turn, its Bianchi identity is obtained from (\ref{Bianchi}) as
\begin{eqnarray}
D^{\vphantom{(}}_{[\mu} {\cal H}^{(4)}_{\nu\rho\sigma\tau]\, r} &=&
-2\,b_{I rs}\,{\cal F}_{[\mu\nu}^s\,{\cal H}_{\rho\sigma\tau]}^I
+ \dots
\;,
\label{Bianchi2}
\end{eqnarray}
where the ellipsis represents possible terms that
vanish under projection with $g^{Ir}$.
We finally note that the \emph{general} variation of the field-strengths is given by
\begin{eqnarray}
\delta {\cal F}_{\mu\nu}^r &=& 2 D_{[\mu} \delta A_{\nu]}^r + h^r{}_I\,\Delta B_{\mu\nu}^I
\;,\nonumber\\[.5ex]
\delta {\cal H}_{\mu\nu\rho}^I &=& 3 D_{[\mu} \Delta B_{\nu\rho]}^I
+6\, d^I{}_{rs} \, {\cal F}_{[\mu\nu}^r \,\delta A_{\rho]}^s + g^{Ir}\,\Delta C_{\mu\nu\rho\,r}
\;,\nonumber\\[.5ex]
\delta {\cal H}^{(4)}_{\mu\nu\rho\sigma\,r} &=& 4 D_{[\mu} \Delta C_{\nu\rho\sigma]r}
-6\, b_{Irs}\,{\cal F}_{[\mu\nu}^s\,\Delta B_{\rho\sigma]}^I
+ 4\, b_{Irs}\,{\cal H}_{[\mu\nu\rho}^I\,\delta A_{\sigma]}^s + \dots
\;,\label{Delta2}
\end{eqnarray}
again with the ellipsis representing possible terms that
vanish under projection with~$g^{Ir}$.

This vector/tensor gauge system is completely defined by the choice of the
invariant tensors $g^{Ir}$, $h^r{}_I$, $b_{Irs}$, $d^I{}_{rs}$, and $f_{rs}{}^t$.
Consistency of the tensor hierarchy, i.e.\ covariance of the field strengths (\ref{defF})
requires that the gauge group generators
in the various representations are given by
\begin{eqnarray}
(X_{r})_{s}{}^t &=&(X^{\rm V}_{r})_{s}{}^t~\equiv~ -f_{rs}{}^t + h^t{}_I\,d^I{}_{rs}
\;,\nonumber\\
(X_{r})_{I}{}^J &=& (X^{\rm T}_{r})_{I}{}^J~\equiv~ 2\, d^J{}_{rs} h^s{}_I -g^{Js} b_{Isr}
\;,
\label{genpar}
\end{eqnarray}
in terms of the invariant tensors parametrizing the system.
Further constraints follow from closure of the algebra (or generalized Jacobi identities)
\begin{eqnarray}
[X_r, X_s] &=& -(X_{r})_{s}{}^t \,X_t
\;,
\label{jacobi}
\end{eqnarray}
and gauge invariance of the tensors $d^I{}_{rs}$ and $b_{I\,rs}$:
\begin{eqnarray}
2(X_r)_{(p}{}^s\,d^I{}_{q)s} - (X_r)_J{}^I\,d^J{}_{pq} &=& 0
\;,\label{i1}
\\
(X_r)_p{}^s\,b_{I\,sq}+ (X_r)_q{}^s\,b_{I\,ps}+(X_r)_I{}^J\,b_{J\,pq} &=& 0
\;.
\label{i2}
\end{eqnarray}
Using (\ref{genpar}), (\ref{i1}) and (\ref{i2}) the generalized Jacobi identities (\ref{jacobi}) take the equivalent form
\begin{eqnarray}
{}
h^r{}_I g^{Is} &=& 0
\;,\label{c1}
\\[.4ex]
{}
f_{rs}{}^t h^r{}_I - d^J{}_{rs}\,h^t{}_J h^r{}_I &=& 0
\;,\label{c2}
\\[.4ex]
{}
f_{[pq}{}^u f_{r]u}{}^s - \frac13 h^s{}_I\, d^{I}{}_{u[p} f_{qr]}{}^u &=& 0
\;,\label{c3}
\\[.4ex]
{}
g^{Js} h^r{}_K b_{Isr} - 2h^s{}_I h^r{}_K \, d^J{}_{rs} &=& 0
\;,\label{c4}
\\[.4ex]
{}
-f_{rt}{}^s g^{It} + d^J{}_{rt} h^s{}_J g^{It} - g^{It} g^{Js} b_{Jtr} &=& 0
\;.
\label{c5}
\end{eqnarray}

\subsection{Supersymmetry and field equations}
\label{subsec:susy}

It has been shown in~\cite{Samtleben:2011fj} that the supersymmetrization of the
non-abelian vector/tensor gauge system is essentially unique and determines the equations
of motion\footnote{Generically certain projections of the Yang--Mills fields may remain off-shell.}. According to ${\cal N}=(1,0)$ supersymmetry, the vector fields are
embedded into off-shell vector multiplets $\{A^r, \lambda^{i\,r}, Y^{ij} \}$, while the tensor fields
become part of an on-shell self-dual tensor multiplet $\{B_{\mu\nu}^I, \chi^{i\,I}, \phi^I\}$.
The three-form tensor fields $C_{\mu\nu\rho\,r}$ do not represent new dynamical degrees
of freedom and do not introduce additional superpartners.
The general supersymmetry transformation rules are
\begin{eqnarray}
\delta A^r_\mu&=& -\bar\epsilon\gamma_\mu\lambda^r\;,\nonumber\\[.5ex]
\delta \lambda^{i\,r} &=& \frac18\,\gamma^{\mu\nu} {\cal F}^r_{\mu\nu} \epsilon^i
-\frac12\,Y^{ij\,r}\epsilon_j + \frac14 h^r{}_I \phi^I \epsilon^i \;,\nonumber\\[.5ex]
\delta Y^{ij\,r} &=& - {\bar\epsilon}^{(i}\gamma^\mu D_\mu\lambda^{j)r}
+2 h^r{}_I\,\bar\epsilon^{(i} \chi^{j) I}
\;,\nonumber\\
\delta \phi^I &=& \bar\epsilon\chi^I\;,\nonumber\\
\delta\chi^{i\,I} &=& \frac1{48} \, \gamma^{\mu\nu\rho} \,{\cal H}^{I}_{\mu\nu\rho}
\epsilon^i +\frac14\,\gamma^\mu D_\mu\phi^I \epsilon^i
- \frac12 d^I{}_{rs} \gamma^\mu\lambda^{i\,r}\,
\bar\epsilon\gamma_\mu \lambda^s\;,
\nonumber\\
\Delta B^I_{\mu\nu} &=& -\bar\epsilon\gamma_{\mu\nu}\chi^I\;,
\nonumber\\[1ex]
g^{Ir} \Delta C_{\mu\nu\rho\,r} &=& - g^{Ir} b_{Jrs} \, \bar\epsilon\gamma_{\mu\nu\rho}\lambda^s \phi^J
\;.\label{tensusy}
\end{eqnarray}
They are obtained from the free field transformation rules by covariantizing and
adding terms proportional to $h^r{}_I$ etc.\ such that the supersymmetry algebra closes
on the fields of the tensor multiplet, provided these fields in addition satisfy the field equations
\begin{eqnarray}
{\cal H}^{I\,-}_{\mu\nu\rho} &=&-  d^I{}_{rs} \bar\lambda^r \gamma_{\mu\nu\rho} \lambda^s
\;,
\nonumber\\[1ex]
\gamma^\sigma D_\sigma \chi^{iI} &=&
\frac12  d^I{}_{rs} {\cal F}^r_{\sigma\tau}\, \gamma^{\sigma\tau}\lambda^{is}
+  2 d^I{}_{rs} Y^{ij\,r}\, \lambda^{s}_j
+\left(d^I{}_{rs}h^s{}_J - 2 b_{J sr}g^{Is}  \right) \phi^J \lambda^{ir}
\;,
\nonumber\\[1ex]
D^\mu D_\mu\,\phi^I &=&
-\frac12d^I{}_{rs} \left(
 {\cal F}_{\mu\nu}^r {\cal F}^{\mu\nu\, s} -4\,Y_{ij}^{r} Y^{ij\,s} + 8 \bar\lambda^r \gamma^\mu D_\mu \lambda^s
\right)
\nonumber\\[.4ex]
&&{}
-2 \left(b_{J sr}g^{Is} -8 d^I{}_{rs}h^s{}_J  \right) \bar\lambda^r \chi^J
-3\, d^I{}_{rs}h^r{}_J h^s{}_K\,
\phi^J\phi^K
\;.
\label{eomten}
\end{eqnarray}
The minimal conditions for supersymmetry to close on these equations are given by the following equations for the
Yang-Mills multiplet:
\begin{eqnarray}
g^{Kr}b_{Irs}\,\left( Y^s_{ij} \,\phi^I - 2 \bar\lambda^s_{(i} \chi_{j)}^I \right) &=& 0
\;,
\nonumber\\[1ex]
g^{Kr}b_{Irs}\left({\cal F}_{\mu\nu}^s \phi^I-2\,\bar\lambda^s \gamma_{\mu\nu} \chi^I \right)
&=& \tfrac{1}{4!}\,\varepsilon_{\mu\nu\lambda\rho\sigma\tau}\,g^{Kr}\,{\cal H}^{(4)\,\lambda\rho\sigma\tau}_{r}
\;,
\nonumber\\[1ex]
g^{Kr}b_{Irs} \left(
\phi^I \gamma^\mu D_{\mu} \lambda_i^{s}
+\tfrac{1}{2} \gamma^\mu  \lambda_i^{s} D_{\mu}\phi^I
\right) &=&
g^{Kr}b_{Irs} \left(
\tfrac{1}{4}  {\cal F}_{\mu\nu}^s\gamma^{\mu\nu}  \chi_i^{I}
+\tfrac{1}{24} {\cal H}_{\mu\nu\rho}^I\gamma^{\mu\nu\rho}  \lambda_i^{s}
- Y_{ij}^s\,\chi^{j\,I}
\right.
\nonumber\\[.5ex]
&&{}
\qquad\;\;\left.{}
+ \tfrac{3}{2} h^s_J \phi^I \chi_i^{J}
+\tfrac{1}{3} d^I_{uv}\,\gamma^\mu \lambda_i^{u} \bar\lambda^s \gamma_\mu \lambda^v
\right)
\;.
\label{eomYM}
\end{eqnarray}
These are not standard field equations in the sense that they generically induce further constraints on
the fields, as we will discuss below for specific examples. In addition certain projections of the auxiliary
Yang--Mills field $Y^{ij}$ are undetermined by these equations such that the corresponding multiplet
remains off-shell.

A stronger version of these equations is given by the above equations without the contraction with the tensor $g^{Kr}$. However,
in order that these equations close under supersymmetry the inclusion of a four-form gauge potential is necessary. We refer
to \cite{Samtleben:2011fj} for the details. Next we discuss models which provide an action, in a certain sense,
in which case one encounters equations that are not contracted with the $g^{Kr}$ tensor.

\subsection{Models admitting an action}
\label{subsec:action}

The field equations described above can be derived from an action (modulo the standard subtleties
for the construction of six-dimensional actions for self-dual tensor fields, cf.\ \cite{Henneaux:1988gg,Pasti:1997gx}),
if the parameters satisfy a number of further constraints:
\begin{eqnarray}
h^r{}_I &=& \eta_{IJ}g^{Jr}
\;,\qquad
b_{I\,rs} ~=~ 2\eta_{IJ}d^J{}_{rs}
\;,
\qquad
\eta_{IJ}\,d^I{}_{p(q} d^J{}_{rs)} ~=~0
\;,
\label{conaction}
\end{eqnarray}
with a non-degenerate symmetric metric $\eta_{IJ}$\,.
The Lagrangian is given by
\begin{eqnarray}
{\cal L} &=&
-\frac18 D^\mu \phi_I \,D_\mu \phi^I
-\frac12 \bar\chi_I\, \gamma^\mu D_\mu \chi^{I}
+\frac1{16} b_{I rs} \phi^I \left(
 {\cal F}_{\mu\nu}^r {\cal F}^{\mu\nu\, s}
-4 Y_{ij}^{r} Y^{ij\,s} +8  \bar\lambda^r \gamma^\mu D_{\mu} \lambda^{s} \right)
\nonumber\\[.5ex]
&&{}
-\frac1{96}  {\cal H}_{\mu\nu\rho}^I\, {\cal H}^{\mu\nu\rho}_I
-\frac1{48} b_{Irs} {\cal H}_{\mu\nu\rho}^I\,\bar\lambda^r\gamma^{\mu\nu\rho}  \lambda^{s}
- \frac14 b_{Irs} {\cal F}_{\mu\nu}^r\,\bar\lambda^s\gamma^{\mu\nu}  \chi^{I}
+ b_{Irs} Y_{ij}^r\,\bar\lambda^{i\,s}\chi^{j\,I}
\nonumber\\[.4ex]
&&{}
+  \frac12\left( b_{J sr} g_I{}^{s}  -4  b_{I sr} g_J{}^{s} \right) \phi^I  \bar\lambda^r \chi^J
+  \frac18  b_{Irs} g_J{}^r g_K{}^s \,\phi^I \phi^J\phi^K
-\frac1{48} {\cal L}_{\rm top}
\nonumber\\[.5ex]
&&{}
-\frac1{24} b_{Irs} b^I{}_{uv}\,\bar\lambda^r\gamma^\mu \lambda^{u} \bar\lambda^s \gamma_\mu \lambda^v
\;,
\label{action}
\end{eqnarray}
where the topological term $\mathcal L_{\rm top}$ was described in \cite{Samtleben:2011fj}. For the tensor multiplet, this action has to be supplemented with the first-order self-duality equation
\begin{eqnarray}
{\cal H}^{I\,-}_{\mu\nu\rho} &=&-  d^I_{rs} \bar\lambda^r \gamma_{\mu\nu\rho} \lambda^s
\;,
\label{firsto}
\end{eqnarray}
to be imposed {\em after} having derived the second-order equations of motion, just as in the democratic formulation
of ten-dimensional supergravities~\cite{Bergshoeff:2001pv}.\footnote{
Alternatively, this self-duality can be implemented by using a non-abelian
version~\cite{Samtleben:2011eb} of the Henneaux-Teitelboim action~\cite{Henneaux:1988gg}
that breaks manifest space-time covariance.}

For the Yang-Mills multiplet the equations of motion take the form of (\ref{eomYM}), but the first and the last
equation occur \emph{without} the contraction with
the $g^{Kr}$ tensor. In addition, from the variation w.r.t.\ the gauge field one obtains the second-order equation
\begin{eqnarray}
  \label{YM}
  b_{Irs} \,D^\nu \left( \phi^I {\cal F}_{\mu\nu}^s - 2 \bar\lambda^s\gamma_{\mu\nu}\chi^I \right)
&=&
\left(\phi^ID_\mu \phi^J-2 \bar\chi^I \gamma_\mu \chi^J\right) X_{r\,IJ}
-2 \phi^I b_{I pq} X_{rs}{}^q\, \bar\lambda^p \gamma_\mu \lambda^s
\nonumber\\[.5ex]
&&{}
-\tfrac{1}{12}b_{Irs}\, \varepsilon_{\mu\nu\rho\lambda\sigma\tau}\,
{\cal F}^{\nu\rho\,s} {\cal H}^{\lambda\sigma\tau\,I}
\;,
\end{eqnarray}
which can be obtained as the covariant derivative of the uncontracted duality equation for the two-form and four-form
field strength in (\ref{eomYM}), see \cite{Samtleben:2011fj} for more details.

The scalar kinetic term in the action (\ref{action}) turns out not to be positive-definite. In order to
see if ghosts might appear as physical states and destroy unitarity a complete analysis of the
huge (tensor) gauge symmetry and the associated constraints is necessary. In addition, the equations
of motions can imply further constraints and eliminate (ghost) degrees of freedom. A priori the action contains
also a cubic and thus unbounded scalar potential, if the multiplying invariant tensor structure exists
for the considered group and representations. Such a non-positive potential is consistent with supersymmetry exactly because
the model might have ghost states. For the case that the potential does not exist, its vanishing can be taken as
an indication that it might be possible to remove the ghost states consistently from the physical spectrum.

The kinetic matrix of the Yang-Mills multiplet,
$K_{rs}\equiv \phi^I b_{Irs}$, shows that the Lagrangian definition of the models is useful only in the
conformal broken phase, where the scalar of the tensor multiplet assumes a non-vanishing vev, which acts as an inverse Yang-Mills coupling.
However, the kinetic matrix $\langle K_{rs} \rangle = \langle \phi^I\rangle b_{Irs}$ may have null directions in
which requires further analysis.


Similar structures as the couplings of (\ref{action})
have appeared in generic 6d supergravity theories and the discussion of conformal fixed points in six-dimensional gauge theories
\cite{Nishino:1986dc,Sagnotti:1992qw,Ferrara:1997gh,Ferrara:1996wv,Nishino:1997ff,Riccioni:2001bg,Seiberg:1996qx}.

We conclude with a presentation of the superconformal symmetry transformations \cite{Sezgin:1994th}.
Denoting the fields in the theory by $\Phi=(\phi^I, B_{\mu\nu}^I,\chi^I,A_\mu^r, Y^{ij},\lambda^r, C_{\mu\nu\rho r})$,
the conformal transformations are given by
\begin{equation}
  \label{eq:2}
  \delta_C \Phi = {\cal L}_\xi \Phi  + \lambda_D \Omega \Phi\ ,
\end{equation}
where ${\cal L}_\xi$ is the Lie derivative with respect to the conformal Killing vector defined by
$\partial_{(\mu} \xi_{\nu)}=\Omega\eta_{\mu\nu}$, which also defines $\Omega$,
and $\lambda_D$ is the Weyl weight for $\Phi$ given by $(2,0,5/2,0,2,3/2,0)$.
The Lie derivative for the fermionic fields $\Psi=(\chi^I, \lambda^r)$, in particular, takes the form
${\cal L}_\xi \Psi = \xi^\mu\partial_\mu\Psi +\frac14 \partial_\mu\xi_\nu \gamma^{\mu\nu}\Psi$. The conformal
supersymmetry transformations, on the other hand, involve conformal Killing spinors, consisting of a pair $(\eta_+, \eta_-)$
that satisfy $\partial_\mu\eta_+ -\frac12 \gamma_\mu\eta_-=0$. The superconformal transformations take the form of supersymmetry
transformations in which the constant supersymmetry parameter $\epsilon$ is replaced by $\eta_+$, and the parameter $\eta_-$ arises
only in $\delta_{\eta_-} \chi^I = -\frac12 \phi^I \eta_-$. Note that the bosonic conformal transformation together with supersymmetry
ensures the full superconformal symmetry since the commutator of a conformal boost with supersymmetry yields the special supersymmetry
generator \cite{Bergshoeff:1985mz}.

\section{Solving the generalized Jacobi identities}
\label{subsec:solving}


Consistency of the non-abelian vector/tensor gauge system is encoded in the set of constraints
(\ref{i1}) -- (\ref{c5}) which generalize the standard Jacobi identity of pure Yang-Mills theory. At first sight
these equations appear quite intractable, but some non-trivial solutions have been found in~\cite{Samtleben:2011fj,Chu:2011fd}.
Several of them are based on the group-theoretical considerations of gauged supergravity theories
in six dimensions which can be employed to derive particular solutions to the system of constraints.\footnote{
This does not necessarily imply that the resulting superconformal models can be embedded into supergravity.}
In this section, we will reduce the system (\ref{i1}) -- (\ref{c5}) by an explicit choice of basis which
allows to exhibit the underlying structure and construct a rather large class of solutions.

For an arbitrary matrix $h^r{}_I$, we can
choose a basis in the space of vector and two-form tensor fields according to
a split $A_\mu^r \longrightarrow \{A_\mu^\alpha , A_\mu^a\}$ and
$B_{\mu\nu}{}^I \longrightarrow \{B_{\mu\nu}{}_{a'} , B_{\mu\nu}{}^a\}$,
such that the matrix $h^r{}_I$ takes diagonal form
\begin{eqnarray}
h^r{}_I &=&
 \left(
\begin{array}{cc}
h^\alpha{}^{b'} & h^\alpha{}_b \\
h^{a}{}^{b'} & h^a{}_b
\end{array}
\right)
~=~
\left(
\begin{array}{cc}
0 & 0 \\
0 & \delta^a_b
\end{array}
\right)\;,
\label{r1}
\end{eqnarray}
with indices $a=1, \dots, {\rm rk}\,h$\, and indices $\alpha$, and $a'$
labeling the complement of the spaces of vector and tensor fields, respectively.
In this basis (\ref{c1}) is solved by
\begin{eqnarray}
g^{Ir} &=&
 \left(
\begin{array}{cc}
g_{a'}{}^\alpha & g_{a'}{}^b \\
g^{a}{}^\alpha & g^a{}^b
\end{array}
\right)
~=~
\left(
\begin{array}{cc}
g_{a'}{}^\alpha & g_{a'}{}^b \\
0 & 0
\end{array}
\right)\;,
\label{r2}
\end{eqnarray}
with arbitrary blocks $g_{a'}{}^\alpha$, $g_{a'}{}^b$.
In this basis the constraints (\ref{c2}) -- (\ref{c3}) translate as follows:
the components $f_{\alpha\beta}{}^\gamma$ are the structure constants of a
Lie algebra $\mathfrak{g}$, satisfying standard Jacobi identities. Moreover, we find that
\begin{eqnarray}
 f_{ab}{}^c &=& 0\ ,\quad f_{ab}{}^\alpha=0\ ,\quad
f_{a\alpha}{}^\beta = 0\ ,\qquad f_{\alpha a}{}^b =-\frac12\,(T_\alpha)_a{}^b\ ,
\label{r3}
\end{eqnarray}
and
\begin{eqnarray}
d^c{}_{ab} &=& 0\ ,\qquad d^b{}_{\alpha a} = \frac12\,(T_\alpha)_a{}^b\ ,\quad
\label{r4}
\end{eqnarray}
where $T_\alpha$ are the generators of the Lie algebra $\mathfrak{g}$
in some representation ${\cal R}$, i.e.\
\begin{eqnarray}
[T_\alpha, T_\beta] &=& f_{\alpha\beta}{}^\gamma \,T_\gamma
\;,
\end{eqnarray}
with dimension ${\rm dim} \,{\cal R}={\rm rk}\,h$\,.
From (\ref{c4}), it further follows that
\begin{eqnarray}
g_{a'}{}^r b^{b'}{}_{ra} &=& 0\;,\qquad
2d_{c'\,}{}_{ab} ~=~ g_{c'}{}^r\,b_{b\,ra}
\;.
\label{cc4}
\end{eqnarray}
while equation (\ref{c5}) states that the matrix $g^{Ir}$ is gauge invariant under
the action of the generators (\ref{genpar}).
Together, we deduce that the generators (\ref{genpar}) take the form
\begin{eqnarray}
X_\alpha^{\rm V} &=&
 \left(
\begin{array}{cc}
-f_{\alpha\beta}{}^\gamma &\;\;\; -f_{\alpha\beta}{}^b +d^b{}_{\alpha\beta} \\
0 & (T_\alpha)_a{}^b
\end{array}
\right)
\;, \qquad X_a^{\rm V} ~=~ 0
\;,
\label{genV}
\end{eqnarray}
in the vector sector, and
\begin{eqnarray}
X_\alpha^{\rm T} &=&
 \left(
\begin{array}{cc}
 -g_{b'}{}^r b^{a'}{}_{r\alpha} & 0  \\
2d_{b'}{}_{\alpha a}-g_{b'}{}^r b_{a\,r\alpha} &(T_\alpha)_a{}^b
\end{array}
\right)
\;, \qquad X_a^{\rm T} ~=~ 0
\;,
\label{genT}
\end{eqnarray}
in the tensor sector. The Jacobi identity (\ref{jacobi}) reduces to
\begin{eqnarray}
[X_\alpha, X_\beta] &=& f_{\alpha\beta}{}^\gamma \,X_\gamma
\;.
\end{eqnarray}
To summarize, we have reduced the original system of constraints (\ref{i1}) -- (\ref{c5}) by an explicit choice of
basis without any loss of generality and the result is given by (\ref{r1}) -- (\ref{cc4}), where all non-vanishing tensors are invariant under the action of the generators (\ref{genV}) and (\ref{genT}), forming the Lie algebra $\mathfrak{g}$  with structure constants $f_{\alpha\beta}{}^\gamma$.


\section{The class based on semi-simple Lie algebras}
\label{sec:123}


Under certain additional assumptions, we can explicitly solve the remaining constraints and obtain a large class of solutions. Specifically, we will restrict to the case of a {\em semi-simple} Lie algebra ${\mathfrak{g}}$ and
{\em non-trivial} representations $T_\alpha$.
In this case, by proper choice of basis,
the matrices $X_\alpha$ from (\ref{genV}), (\ref{genT}) can be taken to be block-diagonal, i.e.\
\begin{eqnarray}
f_{\alpha\beta}{}^b &=&0 ~=~d^b{}_{\alpha\beta}\ ,\qquad
2d_{b'}{}_{\alpha a}~=~g_{b'}{}^r b_{a\,r\alpha}\ ,
\label{css}
\end{eqnarray}
and the generators take the form
\begin{eqnarray}
X_\alpha^{\rm V} &=&
 \left(
\begin{array}{cc}
-f_{\alpha\beta}{}^\gamma & 0 \\
0 & (T_\alpha)_a{}^b
\end{array}
\right)
\;,
\qquad
X_\alpha^{\rm T} ~=~
 \left(
\begin{array}{cc}
 (T'_\alpha)^{a'}{}_{b'}& 0 \\
0 & (T_\alpha)_a{}^b
\end{array}
\right)
\end{eqnarray}
with the matrices $ (T'_\alpha)^{a'}{}_{b'}= -g_{b'}{}^r b^{a'}{}_{r\alpha}$ describing another representation ${\cal R}'$ of $\mathfrak{g}$\,.
The representation ${\cal R}'$ can be only of a few different types:
According to (\ref{r2}), a non-vanishing $g^{Ir}$ requires the representation ${\cal R}'$ to contain
the adjoint representation of $\mathfrak{g}$ or the representation contragredient to ${\cal R}$. For vanishing $g^{Ir}$
on the other hand, the form of (\ref{genT}) shows that the representation ${\cal R}'$ is trivial.
In the following, we discuss separately these three cases.


\subsection{Type I: ${\cal R}'$ is trivial}
\label{subsec:typeI}


This is the case, the matrix $g^{Ir}$ vanishes identically and according to (\ref{conaction}) the corresponding model does not
admit an action. In addition, there are no St\"uckelberg-type couplings among two- and three-form tensor fields, and the
constants $b_{I\,rs}$ are $\mathfrak{g}$ invariant but otherwise unconstrained. To define a minimal model in this category
we shall thus set $b_{I\,rs}=0$. In this case the only non-vanishing components of the $g,h,f,d$ tensors are
\begin{eqnarray}
 f_{\alpha\beta}{}^\gamma\ ,\quad h^a{}_b=\delta^a_b\ , \quad
 f_{\alpha a}{}^b =-\frac12\,(T_\alpha)_a{}^b\ ,\quad
 d^b{}_{\alpha a} = \frac12\,(T_\alpha)_a{}^b\ ,\quad
d_{c'\alpha\beta}=d_{c'}\eta_{\alpha\beta}\ ,
\label{a1}
\end{eqnarray}
where $f_{\alpha\beta}{}^\gamma$ and $\eta_{\alpha\beta}$ are the structure constants
and Cartan-Killing form of a semi-simple Lie algebra $\mathfrak{g}$, respectively,
and $d_{c'}$ are arbitrary constants.
As the 3-form potential ${\cal C}_{\mu\nu\rho r}$ does not couple to the rest of the system, we shall set it to zero. The resulting model has the fields
\begin{eqnarray}
(A_\mu^\alpha, \lambda^{i\alpha},Y^{ij \alpha})\ ,\quad
(A_\mu^a, \lambda^{ia}, Y^{ij a}, B_{\mu\nu}^a, \chi^{ia}, \phi^a)\ ,
\label{fc1}
\end{eqnarray}
and a set of gauge singlet tensor multiplet fields $(B_{\mu\nu a'}, \chi^i_{a'}, \phi_{a'})$ where the indices $(\alpha, a,a')$ label  the adjoint representation, an irreducible representation ${\cal R}$ and singlets of $\mathfrak{g}$, respectively. The field equations for the singlet tensor multiplet of fields are precisely those described long ago in~\cite{Bergshoeff:1996qm}, and as they do not involve the fields of the tensor-YM multiplet, we shall leave them out and refer to the resulting model as Type I.

The explicit bosonic field strengths arising in the Type I model are\footnote{
In the particular case, when ${\cal R}$ is the adjoint representation, this gauge structure was considered in~\cite{Chu:2011fd}.}
\begin{eqnarray}
{\cal F}_{\mu\nu}^\alpha &=&
2 \partial_{[\mu} A_{\nu]}^\alpha - f_{\beta\gamma}{}^\alpha A_\mu^\beta A_\nu^\gamma ~\equiv~ { F}_{\mu\nu}^\alpha
\;,\nonumber\\[.5ex]
{\cal F}_{\mu\nu}^a &=&  2 D_{[\mu} A_{\nu]}^a -  (T_{\alpha})_{b}{}^a A_{[\mu}^\alpha A_{\nu]}^b
+ B_{\mu\nu}^a ~\equiv~ {\cal B}_{\mu\nu}^a\ ,
\nonumber\\[.5ex]
{\cal H}_{\mu\nu\rho}^a &=& 3 D_{[\mu} {\cal B}_{\nu\rho]}^a\,.
\label{FBH}
\end{eqnarray}
The supersymmetry transformations of the off-shell YM multiplet are
\begin{eqnarray}
\delta A^\alpha_\mu&=& -\bar\epsilon\gamma_\mu\lambda^\alpha\;,\nonumber\\[.5ex]
\delta \lambda^{i\,\alpha} &=& \frac18\,\gamma^{\mu\nu} { F}^\alpha_{\mu\nu} \epsilon^i
-\frac12\,Y^{ij\,\alpha}\epsilon_j \ ,
\nonumber\\[.5ex]
\delta Y^{ij\,\alpha} &=& - {\bar\epsilon}^{(i}\gamma^\mu D_\mu\lambda^{j)\alpha}\ ,
\label{SYM}
\end{eqnarray}
and those of the tensor-YM multiplet are
\begin{eqnarray}
\delta \phi^a &=& \bar\epsilon\chi^a\;,\nonumber\\[.5ex]
\delta {\cal B}_{\mu\nu}^a &=& 2 \, \bar\epsilon\,\gamma_{[\mu}D_{\nu]} \lambda^a -\bar\epsilon\gamma_{\mu\nu}\chi^a
\;,\nonumber\\[.5ex]
\delta \lambda^{i\,a} &=& \frac18\,\gamma^{\mu\nu} {\cal B}^a_{\mu\nu} \epsilon^i
-\frac12\,Y^{ij\,a}\epsilon_j + \frac14  \phi^a \epsilon^i \;,\nonumber\\[.5ex]
\delta\chi^{i\,a} &=& \frac1{48}  \gamma^{\mu\nu\rho}\, {\cal H}_{\mu\nu\rho}^a\,
\epsilon^i +\frac14\,\gamma^\mu D_\mu\phi^a \epsilon^i
- \frac12 (T_{\alpha})_b{}^a\, \gamma^\mu\lambda^{i\,(\alpha}\, \bar\epsilon\gamma_\mu \lambda^{b)}
\;,
\nonumber\\[.5ex]
\delta Y^{ij\,a} &=& - {\bar\epsilon}^{(i}\gamma^\mu D_\mu\lambda^{j)a}
+2 \,\bar\epsilon^{(i} \chi^{j) a}\,.
\label{ST}
\end{eqnarray}
The field equations for this model are
\begin{eqnarray}
{\cal H}^{a\,-}_{\mu\nu\rho} &=& -(T_{\alpha})_b{}^a\, \bar\lambda^b \gamma_{\mu\nu\rho} \lambda^\alpha
\;,
\nonumber\\[1ex]
\slashed{D} \chi^{ia} &=&
(T_{\alpha})_b{}^a \left(\frac14 \gamma^{\mu\nu} {\cal B}_{\mu\nu}^b\lambda^{i\alpha}
+\frac14\gamma^{\mu\nu} { F}_{\mu\nu}^\alpha\lambda^{ib}
+2 Y^{ij(b}\, \lambda^{\alpha)}_j +\frac12\phi^b \lambda^{i\alpha}\right)
\;,
\nonumber\\[1ex]
D^\mu D_\mu\,\phi^a &=&  (T_{\alpha})_b{}^a
 \left(
 -\frac12{\cal B}_{\mu\nu}^b { F}^{\mu\nu\, \alpha}
  +2\,Y_{ij}^b Y^{ij\,\alpha} -4 \bar\lambda^{(b}\slashed{D} \lambda^{\alpha)}
 +8 \bar\lambda^\alpha\chi^b \right)\,.
\label{eomI}
\end{eqnarray}
With the described choices for the invariant tensors this system closes under supersymmetry without any
on-shell condition for the vector multiplets, see (\ref{eomYM}).
Hence, the Type I model does not determine the complete dynamics of the system and either supersymmetric equations
of motions for the Yang-Mills multiplets should be imposed, or they should be treated as background fields.
However, for groups/representations
which allow for non-vanishing tensors $b_{I\,rs}$ supersymmetric equations for the Yang-Mills multiplets might
be obtained in the context of extended models \cite{Samtleben:2011fj}, given by the uncontracted version of
(\ref{eomYM}). See also the comments below (\ref{eomYM}) and around (\ref{YM}).


\subsection{Type II: ${\cal R}'$ is in the adjoint representation}
\label{subsec:typeII}


Taking ${\cal R}'$ in the adjoint representation implies that
$g_{\beta}{}^\alpha =\delta_{\beta}^\alpha$ and $g_{\beta}{}^b=0$\footnote{The case in which also ${\cal R}$ is the
adjoint representation is included in the type III models described below.}.
In this case the only non-vanishing components of the $g,h,f,d,b$ tensors are\footnote{
For ${\cal R}=0$, this corresponds to the example discussed in section 3.6 of~\cite{Samtleben:2011fj}.}
\begin{eqnarray}
&& h^b{}_a = \delta_a^b\ ,\quad g_\alpha{}^\beta=\delta_\alpha^\beta\ ,
\quad f_{\alpha\beta}{}^\gamma\ ,\quad f_{\alpha a}{}^b =-\frac12\,(T_\alpha)_a{}^b\ ,
\nonumber\\[.5ex]
&& d^b{}_{\alpha a} = \frac12\,(T_\alpha)_a{}^b\ ,\quad d_{\alpha\,\beta\gamma}\ ,
\quad d_{\alpha\,ab}\ , \quad d_{\alpha\,\beta b}\ , \quad b^\alpha{}_{b \gamma}\ , \quad b_{a\, b \gamma}
\nonumber\\[.5ex]
&& b^{\beta}{}_{\gamma\alpha}=f_{\alpha\gamma}{}^\beta\ ,\quad b^\gamma{}_{ab}\ ,\quad
b_{a\,\beta\alpha}=2d_{\beta}{\,}_{\alpha a}\ ,\quad
b_{a\,\alpha b}=2d_{\alpha\,ab}\ ,\quad b_{cab}\ .
\label{r20}
\end{eqnarray}
The undetermined tensors $d_{\alpha\,\beta\gamma}$, $d_{\alpha\,ab}$, $d_{\alpha\,\beta b}$ etc.\
exist only for very particular choices of $\mathfrak{g}$ and ${\cal{R}}$. As the 3-form potential $C_{\mu\nu\rho\,a}$
decouples from the rest of the system we shall set it to zero. Thus, the fields of the model,
which we shall refer to as Type II, are those listed in ({\ref{fc1}) together with
\begin{eqnarray}
(B_{\mu\nu\,\alpha}, \phi_\alpha, \chi^i_\alpha, C_{\mu\nu\rho\,\alpha} )\ .
\label{fc2}
\end{eqnarray}
The field strengths, supersymmetry transformations and field equations for the fields (\ref{fc1}) remain exactly as
in the previous section on the Type I model except that the adjoint Yang--Mills multiplet is no longer off--shell.
The explicit description of the Type II model also requires similar equations for the additional fields listed
in (\ref{fc2}).
These equations can be read off from the general expressions given in (\ref{tensusy}), (\ref{eomten}) and (\ref{eomYM})
by inserting the data summarized in (\ref{r20}). In the simplest case when all undetermined tensors in
(\ref{r20}) vanish the additional supersymmetry transformations become
\begin{eqnarray}
\delta \phi_\alpha&=& \bar\epsilon\chi_\alpha\;,\nonumber\\
\delta\chi^i_\alpha &=& \frac1{48} \, \gamma^{\mu\nu\rho} \,{\cal H}_{\mu\nu\rho\,\alpha}
\epsilon^i +\frac14\,\gamma^\mu D_\mu\phi_\alpha\epsilon^i
\nonumber\\
\Delta B_{\mu\nu\,\alpha} &=& -\bar\epsilon\gamma_{\mu\nu}\chi^I\;,
\nonumber\\
\Delta C_{\mu\nu\rho\,\alpha} &=& - f_{\beta\alpha}{}^\gamma \,\bar\epsilon\gamma_{\mu\nu\rho}\lambda^\beta \phi_\gamma\;,
\end{eqnarray}
while the equations of motion are
\begin{eqnarray}
{\cal H}^{-}_{\mu\nu\rho\,\alpha} =0\;,\qquad\gamma^\sigma D_\sigma \chi^i_\alpha =-2f_{\beta\alpha}{}^\gamma\,\phi_\gamma \lambda^{i\beta}
\;,\qquad D^\mu D_\mu\,\phi_\alpha=-2f_{\beta\alpha}{}^\gamma\bar\lambda^\beta\chi_\gamma
\end{eqnarray}
and
\begin{eqnarray}
f_{\beta\alpha}{}^\gamma\,\left( Y^\beta_{ij} \,\phi_\gamma - 2 \bar\lambda^\beta_{(i} \chi_{\gamma\,j)} \right) &=& 0
\;,
\nonumber\\[1ex]
f_{\beta\alpha}{}^\gamma\left({\cal F}_{\mu\nu}^\beta \phi_\gamma-2\,\bar\lambda^\beta\gamma_{\mu\nu} \chi_\gamma\right)
&=& \tfrac{1}{4!}\,\varepsilon_{\mu\nu\lambda\rho\sigma\tau}\,{\cal H}^{(4)\,\lambda\rho\sigma\tau}_{\alpha}
\;,
\nonumber\\[1ex]
f_{\beta\alpha}{}^\gamma\left(
\phi_\gamma\gamma^\mu D_{\mu} \lambda_i^\beta
+\tfrac{1}{2} \gamma^\mu  \lambda_i^\beta D_{\mu}\phi_\gamma
\right) &=&
f_{\beta\alpha}{}^\gamma\left(
\tfrac{1}{4}  {\cal F}_{\mu\nu}^\beta\gamma^{\mu\nu}  \chi_{\gamma\,i}
+\tfrac{1}{24} {\cal H}_{\mu\nu\rho\,\gamma}\gamma^{\mu\nu\rho}  \lambda_i^\beta
- Y_{ij}^\beta\,\chi^j_\gamma
\right)
\;.
\nonumber\\
\end{eqnarray}
Here we can see explicitly the subtleties in the Yang--Mills sector alluded to earlier. From the first equation it is clear that the auxiliary field $Y^{ij}$ is not completely fixed, it can for example be shifted by a multiple of $\phi$. This implies that some of the Yang--Mills fields remain off--shell. The fact that there are also constraints implied by these equations is easily seen by contracting this equation with $\phi^\alpha$ which gives
\begin{equation}
f_{\beta\alpha}{}^\gamma\,\phi^\alpha\bar\lambda^\beta_{(i} \chi_{\gamma\,j)}= 0\,.
\label{const}
\end{equation}
This constraint eliminates some of the physical degrees of freedom. Such constraints
have to be taken into account when studying the dynamics and degrees of freedom.}


\subsection{Type III: ${\cal R}'$ is in the contragredient representation to ${\cal R}$}
\label{subsec:typeIII}


The remaining case corresponds to taking ${\cal R}'$ in the contragredient representation to ${\cal R}$, and therefore
set $g_{a}{}^\alpha =0$ and $g_{a}{}^b=\delta_a^b$. This is the only case, in which an action is possible
(in the sense discussed above) and we will right away restrict
to the case when the compatibility conditions (\ref{conaction}) are satisfied as well.
In that case the metric
$\eta^{IJ}$ on the tensor sector is given by a matrix with the off-diagonal blocks $\eta_{a}{}^b=\eta^b{}_a =\delta_a^b$.
Taking into account (\ref{conaction}), we deduce that the only nonvanishing components of the tensors $g,h,f,d,b$
are
\begin{eqnarray}
&& h^b{}_a = \delta_a^b\ ,\quad g_a{}^b=\delta_a^b\ ,
\quad f_{\alpha\beta}{}^\gamma\ ,\quad f_{\alpha a}{}^b =-\frac12\,(T_\alpha)_a{}^b\ ,
\nonumber\\[.5ex]
&& d_{abc} = d_{(abc)}\ ,\quad d_{ab\alpha}= d_{(ab)\alpha}\ ,\quad d_{a\beta\gamma}\ ,
\quad d^b{}_{\alpha a}=\frac12 (T_\alpha)_a{}^b\ ,
\nonumber\\[.5ex]
&& b^a{}_{b\alpha}= b^a{}_{\alpha b} = (T_\alpha)_b{}^a\ ,
\quad b_{a\,b\alpha}= b_{a\,\alpha b} = 2d_{(ab)\alpha}\ , \quad b_{b\,ca}=2d_{(abc)}\ , \quad
b_{a\,\beta\gamma}=2d_{a (\beta\gamma)}\, .\nonumber\\
\label{ex3}
\end{eqnarray}
In particular, upon imposing these symmetry properties of $d_{abc}$ and $d_{ab\alpha}$,
the last equation of (\ref{conaction}) is automatically satisfied as a consequence of $\mathfrak{g}$-invariance.
Specifying a model within this class thus corresponds to choosing a Lie algebra $\mathfrak{g}$
with a representation ${\cal R}$, and the $\mathfrak{g}$-invariant tensors
\begin{eqnarray}
 d_{(abc)}\ ,\quad d_{(ab)\alpha}\ ,\quad d_{a(\beta\gamma)}
 \;.
 \label{ddd}
\end{eqnarray}
Again, such tensors only exist for very particular choices of $\mathfrak{g}$ and ${\cal{R}}$\,.
In the following, we will present more explicitly the vector/tensor gauge system
corresponding to this solution of the consistency constraints.
An interesting and straightforward generalization of this solution
would be the inclusion of abelian gauge factors and possible
gauge group singlets among the vector and tensor fields.

As the 3-form potential ${\cal C}_{\mu\nu\rho\,\alpha}$ decouples from the rest of the tensor-YM system, we shall set it to zero. With this understood, we  shall refer to the model defined by (\ref{ex3}) as Type III. Its field content consists of the multiplets listed in (\ref{fc1}) and the additional fields
\begin{eqnarray}
(B_{\mu\nu\, a}, \phi_a, \chi^i_a, C_{\mu\nu\rho\,a} )\ .
\label{fc2b}
\end{eqnarray}
The explicit non-abelian field strengths are those given in (\ref{FBH}) and
\begin{eqnarray}
{\cal H}_{\mu\nu\rho}{}_{\,a} &\equiv& 3 D_{[\mu} B_{\nu\rho]}\,_a  +
6 \, d_a{\,}_{rs}\,  A_{[\mu}^r \partial^{\vphantom{r}}_{\nu\vphantom{]}} A_{\rho]}^s
- 2 f_{pq}{}^s d_a{\,}_{rs}\, A_{[\mu}^r A_{\nu\vphantom{]}}^p A_{\rho]}^q
+C_{\mu\nu\rho\,a} \equiv {\cal C}_{\mu\nu\rho}{}_{\,a}\ ,
\nonumber\\[.5ex]
{\cal H}^{(4)}_{\mu\nu\rho\sigma}{}_{\,a} &\equiv&
4D_{[\mu}^{\vphantom{a}} {\cal C}_{\nu\rho\sigma]\,a}
-6\,d_{a\alpha\beta} {\cal F}^\alpha_{[\mu\nu} {\cal F}^\beta_{\rho\sigma]}
-12\,d_{ab\alpha} {\cal F}^\alpha_{[\mu\nu} {\cal B}^b_{\rho\sigma]}
-6\,d_{abc} {\cal B}^b_{[\mu\nu} {\cal B}^c_{\rho\sigma]}\ .
\label{CH}
\end{eqnarray}
By construction, ${\cal B}^a_{\mu\nu}$ and ${\cal C}_{\mu\nu\rho\,a}$ are invariant under
tensor gauge transformations and transform covariantly under non-abelian vector gauge transformations.
By eliminating $A^a_\mu$ and $B_{\mu\nu\,a}$ from the theory, we have thus
changed the structure of the tensor gauge algebra, similar to choosing the
unitary gauge in the standard Higgs effect.
While this may not be the right formulation for a consistent quantization of the theory,
it allows to illustrate the distribution of degrees of freedom in this model.

Let us finally note that after the redefinitions,
the only appearance of the constant tensors (\ref{ddd}) in the vector/tensor
gauge system, is in the definition of the field strength ${\cal H}^{(4)}$.

The action for the Type III model takes the explicit form
\begin{eqnarray}
{\cal L} &=&
-\frac14 D^\mu \phi_a \,D_\mu \phi^a
- \bar\chi_a\, \gamma^\mu D_\mu \chi^{a}
+\frac1{8} d_{I rs} \phi^I \left(
 {\cal F}_{\mu\nu}^r {\cal F}^{\mu\nu\, s}
-4 Y_{ij}^{r} Y^{ij\,s} +8  \bar\lambda^r \gamma^\mu D_{\mu} \lambda^{s} \right)
\nonumber\\[.5ex]
&&{}
-\frac1{8}  \, {\cal C}^{\mu\nu\rho\,(+)}_a\left(
D^{\vphantom{a}}_\mu {\cal B}_{\nu\rho}^a
+\frac1{3} \,(T_\alpha)_b{}^a \bar\lambda^\alpha\gamma_{\mu\nu\rho}  \lambda^{b}\right)
-\frac1{8} \,d_{ars} \, D^{\vphantom{a}}_\mu {\cal B}_{\nu\rho}^a\,\bar\lambda^r\gamma^{\mu\nu\rho}  \lambda^{s}
\nonumber\\[.5ex]
&&{}
- \frac12 \,d_{Irs} {\cal F}_{\mu\nu}^r\,\bar\lambda^s\gamma^{\mu\nu}  \chi^{I}
+ 2d_{Irs}\, Y_{ij}^r\,\bar\lambda^{i\,s}\chi^{j\,I}
-\frac1{3}  (T_\alpha)_b{}^a \,d_{a\,rs}\,\bar\lambda^r\gamma^\mu \lambda^{\alpha} \bar\lambda^s \gamma_\mu \lambda^b
\nonumber\\[.4ex]
&&{}
+  \frac12\,(T_\alpha)_a{}^b \,\bar\lambda^\alpha \left(\phi^a   \chi_b -4  \phi_b   \chi^a \right)
  -3d_{ab\alpha} \phi^a  \bar\lambda^\alpha \chi^b
- 3d_{abc} \phi^a  \bar\lambda^b \chi^c
+  \frac14  d_{abc}\,\phi^a \phi^b\phi^c
\nonumber\\[.4ex]
&&{}
+\frac1{32}\varepsilon^{\mu\nu\rho\lambda\sigma\tau}
\,  {\cal B}^a_{\mu\nu} \left(
d_{a\alpha\beta}\,   \,{ F}_{\rho\lambda}^\alpha { F}_{\sigma\tau}^\beta
+d_{ab\alpha}\,   \,{ F}_{\rho\lambda}^\alpha {\cal B}_{\sigma\tau}^b
+\frac13d_{abc}\,   \,{\cal B}_{\rho\lambda}^b {\cal B}_{\sigma\tau}^c
\right)
\;.
\label{action_exp}
\end{eqnarray}
Sums over indices $r, s$ still need to be split according to $\lambda^r \rightarrow \{\lambda^\alpha, \lambda^a\}$,
${\cal F}^r \rightarrow \{{ F}^\alpha, {\cal B}^a\}$, etc.

Note that only the self-dual part of the field ${\cal C}_{\mu\nu\rho\,a}$
appears in the Lagrangian and acts as a Lagrange multiplier for the
self-duality equation
\begin{eqnarray}
\left(D^{\vphantom{a}}_{[\mu}{\cal B}^{a}_{\nu\rho]}\right)^- &=&
-  \frac13 (T_\alpha)_b{}^a  \bar\lambda^\alpha \gamma_{\mu\nu\rho} \lambda^b\ ,
\end{eqnarray}
which constitutes half of the additional first-order equations (\ref{firsto}). The other half of (\ref{firsto})
determines the anti-self dual part of ${\cal C}_{\mu\nu\rho\,a}$ to be
\begin{eqnarray}
{\cal C}^{(-)}_{\mu\nu\rho\,a} &=&-  d{}_{ars} \bar\lambda^r \gamma_{\mu\nu\rho} \lambda^s\ .
\label{Cm}
\end{eqnarray}
Since the anti-self-dual part ${\cal C}^{(-)}_{\mu\nu\rho\,a}$
does not show up in the action (\ref{action_exp}), we can impose (\ref{Cm})
as its definition in all previous formulas.
Therefore we find that in contrast to the general Lagrangian (\ref{action}), the
explicit Lagrangian (\ref{action_exp}) for this particular model implies all the
field equations without the need to impose additional first-order duality equations by hand!

In this model, the supersymmetry transformation rules are those given in (\ref{SYM}), (\ref{ST}) and
\begin{eqnarray}
\delta \phi_a &=& \bar\epsilon\chi_a\ ,
\nonumber\\[.5ex]
\delta\chi^{i}_a &=& \frac1{48} \, {\cal C}^{(+)}_{\mu\nu\rho\,a}\,\gamma^{\mu\nu\rho} \,
\epsilon^i +\frac14\,\gamma^\mu D_\mu\phi_a \epsilon^i
- \frac12 d_{ars} \gamma^\mu\lambda^{i\,r}\, \bar\epsilon\gamma_\mu \lambda^s\ ,
\nonumber\\[.5ex]
\delta {\cal C}_{\mu\nu\rho\,a} &=& -3\, \bar\epsilon\,\gamma_{[\mu\nu} D_{\rho]}\chi_a
-6\, d_{ars} \, {\cal F}_{[\mu\nu}^r \,\bar\epsilon\gamma_{\rho]}\lambda^s  - 2d_{Ias} \, \bar\epsilon\gamma_{\mu\nu\rho}\lambda^s \phi^I\ .
\label{tensusy_exp}
\end{eqnarray}
In fact, the relevant part of the last transformation is the projection onto its self-dual part, while
its anti-self dual part just describes the transformation of (\ref{Cm}) into the field equation of $\chi_a$\,.

It is worth noting that the cubic scalar potential of (\ref{action_exp}) contains at most half
of the scalar fields and is entirely triggered by the gauge invariant tensor $d_{abc}$\,.
Moreover, the different $d$-tensors exist only for particular choices of the Lie-algebra $\mathfrak{g}$ and
representations ${\cal R}$.
It is therefore natural (and consistent) to consider generic models where we set all $d$-tensors to zero. In that case
the action simplifies to
\begin{eqnarray}
{\cal L} &=&
-\frac14 D^\mu \phi_a \,D_\mu \phi^a
- \bar\chi_a\, \gamma^\mu D_\mu \chi^a
+\frac1{8}(T_\alpha)_b{}^a\phi_a \left(
 {\cal B}_{\mu\nu}^b{ F}^{\mu\nu\,\alpha}
-4Y_{ij}^b Y^{ij\,\alpha}
+8\bar\lambda^{(b}\gamma^\mu D_{\mu} \lambda^{\alpha)} \right)
\nonumber\\[.5ex]
&&{}
-\frac1{8}{\cal C}^{\mu\nu\rho\,(+)}_a\left(
D^{\vphantom{a}}_\mu {\cal B}_{\nu\rho}^a
+\frac1{3} \,(T_\alpha)_b{}^a \bar\lambda^\alpha\gamma_{\mu\nu\rho}\lambda^b\right)
-\frac14\,(T_\alpha)_b{}^a{\cal B}_{\mu\nu}^b\,\bar\lambda^\alpha\gamma^{\mu\nu}\chi_a
\nonumber\\[.5ex]
&&{}
-\frac14\,(T_\alpha)_b{}^a{ F}_{\mu\nu}^\alpha\,\bar\lambda^b\gamma^{\mu\nu}\chi_a
+2(T_\alpha)_b{}^a\,Y_{ij}^{(b}\,\bar\lambda^{i\,\alpha)}\chi^j_a
+\frac12\,(T_\alpha)_a{}^b \,\bar\lambda^\alpha \left(\phi^a\chi_b-4\phi_b\chi^a\right)
\;.
\end{eqnarray}

The tensor field equations for the multiplet in the representation $\cal R$ are the same as in (\ref{eomI}).
For the tensor multiplet in the contragredient representation $\cal R'$ one obtains,
\begin{eqnarray}
  \label{congred}
  {\cal C}_{\mu\nu\rho\,a}^{(-)} = 0\ , \ \ \slashed{D}\chi^i_a = -2(T_\beta)_a{}^b\bar\lambda^{\beta\,i}\phi_b\ , \ \
   D^\mu D_\mu\phi_a = - 2 (T_\beta)_a{}^b\bar\lambda^{\beta}\chi_b\, .
\end{eqnarray}
This implies that the three-form  ${\cal C}_{\mu\nu\rho\,a}$ may be considered as a self-dual auxiliary form.
In the Yang--Mills sector, for example the auxiliary field equations become
\begin{eqnarray}
(T_\alpha)_a{}^b\left( Y^\alpha_{ij} \,\phi_b - 2 \bar\lambda^\alpha_{(i} \chi_{j)\,b} \right) = 0\;,\quad
(T_\alpha)_a{}^b\left( Y^a_{ij} \,\phi_b - 2 \bar\lambda^a_{(i} \chi_{j)\,b}\right) = 0\; .
\end{eqnarray}
The equations for the Yang-Mills multiplet again imply algebraic conditions on the physical fields, analogous to
(\ref{const}), which might eliminate some of the degrees of freedom. We leave a more careful analysis of these issues
for the future.

Let us finally give the supersymmetry transformation rules in this minimal case.
They are those given in (\ref{SYM}), (\ref{ST}) together with
\begin{eqnarray}
\delta \phi_a &=& \bar\epsilon\chi_a\ ,
\nonumber\\[.5ex]
\delta\chi^{i}_a &=& \frac1{48} \, {\cal C}^{(+)}_{\mu\nu\rho\,a}\,\gamma^{\mu\nu\rho} \,
\epsilon^i +\frac14\,\gamma^\mu D_\mu\phi_a \epsilon^i
\nonumber\\[.5ex]
\delta {\cal C}_{\mu\nu\rho\,a} &=& -3\, \bar\epsilon\,\gamma_{[\mu\nu} D_{\rho]}\chi_a
-(T_\alpha)_a{}^b\,\bar\epsilon\gamma_{\mu\nu\rho}\lambda^\alpha \phi_b\ .
\end{eqnarray}


\section{ Conclusions}
\label{sec:conclusions}


We reviewed the construction of
a general class of six-dimensional (1,0) superconformal models with
non-abelian vector and tensor multiplets as given in \cite{Samtleben:2011fj}.
The formulation of the non-abelian tensor gauge symmetry relies on
a tensor hierarchy for one-, two and non-dynamical three forms.
Here, in particular, we solved for the case of semi-simple groups
the {\em generalized Jacobi identities} (\ref{c5}) of the various invariant tensors that
appear in the tensor hierarchy.

The allowed tensor and vector multiplets are characterized by two
representations ${\cal R}$ and ${\cal R}'$ of the Lie algebra
$\mathfrak{g}$, and its adjoint representation. This corresponds to the decomposition
$r=(\alpha,\, a)$ and $I=(a',\, a)$ of the respective indices, as
given above (\ref{r1}).
For a given representation ${\cal R}$ the various models are classified by the possible
representations ${\cal R}'$. Denoting the vector and tensor multiplets by
${\cal V} = (\, A_\mu,\,\lambda^i,\,Y_{ij}\, )$
and ${\cal T} = (\, B_{\mu\nu},\, \phi,\, \chi^i\, )$ the respective models and their field and representation
content is summarized in table \ref{tab}. The non-dynamical three-form is denoted by $C$.
\\

\begin{table}[!ht]
\label{tab}
{\renewcommand{\arraystretch}{1.5}
\renewcommand{\tabcolsep}{0.2cm}
\begin{tabular}{|c|c|c|c|l|}
\hline
class & ${\cal R}'$ & content & dynamics & parameters\\
\hline
Type I& triv. &$ {\cal V}^\alpha$, $(\, {\cal V}^a,\, {\cal T}^a\,)$, ${\cal T}_{a'}$ & e.o.m.
& $d_{a'}$, $b_{I\, rs}$  \\
Type II & adj. & ${\cal V}^\alpha,\, (\,{\cal T}_\alpha,\, C_\alpha\,)$, $(\, {\cal V}^a,\, {\cal T}^a\,)$ & e.o.m.
& $d_{\alpha\, rs}$,  $b^\alpha{}_{b\gamma},$ \\[-1ex]
 & & &  & $b_{a\, bc},b_{a\, b\gamma},b^\gamma{}_{ab}$\\
Type III &${}^t{\cal R}^{-1}$ & ${\cal V}^\alpha$, $(\,{\cal V}^a,\, {\cal T}^a\,)$, $(\,{\cal T}_{a},\,C_{a}\,)$
& action & $d_{abc},\, d_{a \beta \gamma},\,  d_{a b \gamma}$\\
\hline
\end{tabular}}
\caption{The three types of models for semi-simple groups. The possible representations ${\cal R}'$ are the
trivial, the adjoint and the to ${\cal R}$ contragredient representation.}
\end{table}

Besides the structure constants $f_{\alpha\beta}{}^\gamma$ of 
the Lie algebra $\mathfrak{g}$,
the parameters of the different types of models are the
various undetermined invariant $b$-  and $d$-tensors in the respective
representations. For the type III models the $d$-tensors have to be symmetric in indices of the same kind.
If some of these tensors exist for certain groups and representations they allow for additional interactions
but they can also be consistently set to zero.
An interesting case is given by the type I model, for which the Yang-Mills multiplets are off-shell
for vanishing $b$-tensors. However, in the case of non-vanishing $b$-tensors supersymmetric equations of
motions are determined also for the Yang-Mills multiplet, as discussed in the main text.

For the type III models the equations of motions can be derived from an action,
contrary to the two other classes of models. In all models we
used a field redefinition (or classical gauge) that absorbs the vector in the representation ${\cal R}$
into the two-form, i.e. $(\, A_{\mu}^a,\, B_{\mu\nu}^a\,)\rightarrow {\cal B}_{\mu\nu}^a$, such that
${\cal B}_{\mu\nu}^a$ is invariant under tensor gauge transformations. Similarly, a three-form ${\cal C}_{\mu\nu\rho a}$
invariant under tensor gauge transformations has been defined.

Given the hitherto lack of non-abelian models in six dimensions
the very existence of these models provide new and very intriguing structures
that deserve more study, and it is clear
that there are several open questions
that one still has to address.

The action in the case of the type III models has non-positive definite kinetic terms in the vector and the tensor
sector. It will require further work to understand the fate of the resulting ghost states
and if one can find a mechanism to decouple them. 
In the analysis of physical degrees of freedom one also has to take into account the algebraic constraints
that are implied by the equations of motions of the vector multiplets. 

Perturbatively the models are defined only in the conformal broken phase, where the vev of the
tensor multiplet scalar acts as an inverse Yang-Mills coupling constant.
However, null-directions in the kinetic term for the vector fields implies that the fields of the corresponding vector multiplets become non-dynamical.

It remains to be investigated how much of the presented structures can be carried
over to $(2,0)$ theories. As a first step in this direction,
the inclusion of hypermultiplets to the (1,0) models has been briefly discussed in \cite{Samtleben:2011fj}.
Adding $n_{\rm T}$ hypermultiplets with flat target space completes the present field content from (1,0)
to the (2,0) theories, in addition to the vector multiplets and the three-form potentials.
Although there is no propagating $(2,0)$ vector multiplet, the present construction in (1,0) models
has illustrated the possible relevance of the inclusion of non-propagating degrees of freedom.

Of great relevance is the study of possible M-theory backgrounds for M$5$-branes
that lead to $(1,0)$ supersymmetry and charged tensor multiplets. In general such backgrounds will be
singular spaces, which would also provide propagating vector fields as we have encountered them in our analysis.

A pending question is of course the quantization of the models, and the fate of the conformal symmetry 
at the quantum level. Last but not least, the study of anomalies in the generalized gauge symmetries of the models we have presented here is of great interest.

Finally we want to mention some relations to our results that appeared in the literature.
For example, the analysis of amplitudes for six-dimensional tensor and vector multiplets \cite{Czech:2011dk}
produced a number of interaction terms that we find also in our models and recently a very interesting connection
with seven-dimensional non-abelian Chern-Simons theories has been discussed \cite{Fiorenza:2012tb}.

\bibliographystyle{JHEP2}

\providecommand{\href}[2]{#2}\begingroup\raggedright\endgroup


\end{document}